\documentclass[prb,showpacs,preprintnumbers,amsmath,amssymb]{revtex4}

\usepackage{bm}
\usepackage{dcolumn}
\usepackage[dvipdfm]{graphicx}

\usepackage{color}

\begin{document}

\title{One-dimensional topological insulator: a model for studying finite-size effects in topological insulator thin films}

\author{Mayuko Okamoto}
\author{Yositake Takane}
\author{Ken-Ichiro Imura}

\affiliation{Department of Quantum Matter, AdSM, Hiroshima University, Higashi-Hiroshima 739-8530, Japan}

\date{\today}

\begin{abstract}
As a model for describing finite-size effects in topological insulator thin films,
we study a one-dimensional (1D) effective model of a topological insulator (TI).
Using this effective 1D model,
we reveal the precise correspondence between 
the spatial profile of the surface wave function, and
the dependence of the finite-size energy gap on the thickness ($L_x$) of the film.
We solve the boundary problem both in the semi-infinite and slab geometries
to show that
the $L_x$-dependence of the size gap 
is a direct measure of the amplitude of the surface wave function $\psi (x)$ at the depth of
$x=L_x+1$ [here, the boundary condition is chosen such that $\psi (0)=\bm 0$].
Depending on the parameters,
the edge state function shows either a damped oscillation
(in the ``TI-oscillatory'' region of FIG. \ref{PD}), or becomes overdamped
({\it ibid.}, in the ``TI-overdamped'' phase).
In the original 3D bulk TI,
an asymmetry in the spectrum of valence and conduction bands
is omnipresent.
Here, we demonstrate by tuning this asymmetry one can drive a crossover from
the TI-oscillatory to the TI-overdamped phase.
\end{abstract}


\maketitle

\section{Introduction}

Existence of a protected
\cite{FuKaneMele, FuKane, MooreBalents, Roy}
gapless state is a defining property of the topological insulator, 
which is also considered to be gapless.
\cite{Moore_review, HasanKane, SCZ}
Yet, in realistic systems of a finite size, 
{\it e.g.}, in thin films of a topological insulator, 
\cite{Hirahara, exp_film1, exp_film2, exp_film3, exp_film4}
this last clause is not guaranteed by the topological protection. 
The magnitude of such a finite-size energy gap in the case of a thin film 
is naturally related to the degree of penetration of the surface state wave function 
into the bulk
in the light of the thickness of the film. 
\cite{Shen_PRL, Shen_PRB, Linder, Shen_NJP, Liu_film, weak_eo}
Here, in this paper we point out that this correspondence can be made
more precise.
Indeed, in the system considered
the 
thickness dependence of the finite-size energy gap 
is shown to be a direct measure of the amplitude of the surface wave function.

\section{Reduction of the 3D bulk Hamiltonian to a 1D model}

The physical system we consider is a slab-shaped sample of three-dimensional (3D) topological insulator (TI),
{\it i.e.}, a TI thin film, or a nanofilm.
We first show in this section that finite-size effects in such TI films are described by
an effective 1D model, representing a ``one-dimensional topological insulator''.
Here, let us go back to
the bulk 3D TI, described by the following tight-binding (Wilson-Dirac type) Hamiltonian:
\cite{Liu_PRB, Liu_nphys}
\begin{equation}
h_{3D} (\bm k) = \gamma_0 m(\bm k) + \gamma_\mu t_\mu \sin k_\mu + \epsilon (\bm k) \bm 1_4,
\label{h_3D}
\end{equation}
where
\begin{eqnarray}
m(\bm k) &=& m_0 + 2 \sum_{\mu =x,y,z} m_{2\mu } (1-\cos k_\mu),
\label{mass}
\\
\epsilon (\bm k) &=& \epsilon_0 + 2 \sum_{\mu =x,y,z} \epsilon_{2\mu} (1- \cos k_\mu).
\label{epsilon}
\end{eqnarray}
$\gamma_0$ and three $\gamma_\mu$'s ($\mu = x,y,z$)
are a set of $\gamma$-matrices,
which can be chosen as
\begin{equation}
\gamma_0 = \tau_z,\ 
\gamma_x = \tau_x \sigma_z,\
\gamma_y = \tau_x \sigma_x,\
\gamma_z = \tau_x \sigma_y.
\end{equation}
$\sigma_\mu$'s ($\mu=x,y,z$) are Pauli matrices representing the real spin.
$\bm 1_4$ represents a $4\times 4$ identity matrix, while
the time reversal symmetry requires
$m(\bm k)$ and $\epsilon (\bm k)$ to be an even function of $\bm k$.
The form of the hopping (sine and cosine) terms in Eqs. (\ref{h_3D}), (\ref{mass}) and (\ref{epsilon}) 
reflects that our tight-binding Hamiltonian is implemented on a cubic lattice,
and we consider only nearest neighbor hopping.
Changing the mass parameters: $m_0$ and $m_{2x}$, $m_{2y}$, $m_{2z}$ in Eq. (\ref{mass}),
one can realize different types of weak and strong topological insulating phases.
\cite{weak_eo}

Here, in this paper we study in detail
the spatial profile of the surface wave function in the semi-infinite geometry, and
the finite-size energy gap in the slab geometry,
to reveal a close relation between them.
In the semi-infinite geometry,
the bulk 3D TI occupies the semi-infinite space $x>0$ with a surface on the $(y, z)$-plane.
While, in the slab
the bulk 3D TI is confined to a slab-shaped region $0<x<L_x$.
In the two cases studied, the translational invariance in the $y$- and $z$-directions are
respected, so that $\bm k_\parallel = (k_y, k_z)$ is still a good quantum number.
The energy spectrum is then expressed as
$E=E (\bm k_\parallel)$, 
and $\bm k_\parallel$ belongs to the surface Brillouin zone (BZ).

For keeping the subsequent discussions reasonably simple,
we only consider the case of a slab structure
commensurate with the cubic symmetry; 
here, we place it perpendicular to the $x$-axis ($\bm k_\perp = k_x$).
Then, the cubic symmetry of the lattice and of the BZ ensures
that gapless surface Dirac points appear at the four symmetric points in the surface BZ,
$\bar{\Gamma}=(0,0)$, $\bar{Y}=(\pi,0)$, $\bar{Z}=(0,\pi)$, $\bar{M}=(\pi,\pi)$.
At these symmetric points,
relevant to the low-energy spectrum of the surface states,
the hopping terms in the $y$- and $z$-directions
become inert, {\it i.e.},
\begin{equation}
h_{3D} (k_x, \bm k_\parallel = \bar{\Gamma}, \bar{Y}, \bar{Z}, \bar{M}) 
=\tau_z m(k_x, \bm k_\parallel) 
+ \tau_x \sigma_z t_x \sin k_x,
\end{equation}
where
\begin{equation}
m(k_x, \bm k_\parallel)=\tilde{m}_0 (\bm k_\parallel)+ 2 m_{2x} (1-\cos k_x),
\label{m0_tilde}
\end{equation}
with
\begin{equation}
\tilde{m}_0 (\bm k_\parallel )=\left\{
\begin{array}{ll}
m_0 & (\bm k_\parallel = \bar{\Gamma})\\
m_0+2m_{2y} & (\bm k_\parallel = \bar{Y})\\
m_0+2m_{2z} & (\bm k_\parallel = \bar{Z}) \\
m_0+2(m_{2y}+m_{2z}) & (\bm k_\parallel = \bar{M})
\end{array}
\right..
\label{m0_tilde2}
\end{equation}
There, the $4\times 4$ Hamiltonian matrix can be organized into a block-diagonal form as
\begin{eqnarray}
h_{3D} (k_x, \bm k_\parallel = \bar{\Gamma}, \bar{Y}, \bar{Z}, \bar{M}) &=&
\left[
\begin{array}{c|c}
\begin{array}{cc}
m(k_x, \bm k_\parallel) &  t_x \sin k_x \\
t_x \sin k_x  &  -m(k_x, \bm k_\parallel)  
\end{array}
& \bm 0 \\
\hline
\bm 0 &
\begin{array}{cc}
m(k_x, \bm k_\parallel) &  - t_x \sin k_x \\
- t_x \sin k_x  &  -m(k_x, \bm k_\parallel)
\end{array}
\end{array}
\right]
\nonumber \\
&\equiv&
\left[
\begin{array}{cc}
h_{1D} (k_x) & \bm 0 \\
\bm 0 &  h_{1D} (-k_x)  
\end{array}
\right].
\end{eqnarray}
As we will see explicitly in the following sections,
the surface states of a slab-shaped TI with a finite thickness $L_x$
exhibit indeed a finite-size energy gap,
due to (though this is not the way how we tackle the problem)
overlap of the wave functions on the top and bottom surfaces.
Yet, the magnitude of this energy gap is minimal at
either of these four symmetric $\bm k_\parallel$-points
at which the surface spectrum becomes gapless
in the ideal situation; {\it i.e.}, when $L_x\rightarrow\infty$ or
in a semi-infinite geometry.
Since the real spin down sector ($\sigma_z = -1$) is obtained
from the spin up sector ($\sigma_z = 1$)
by simply reversing the direction of motion,
we can safely focus on the spin up sector
described by the effective 1D theory represented as a $2\times 2$ Hamiltonian matrix
$h_{1D} (k_x)$.

Though this reduction to a $2\times 2$ matrix occurs only at the
four symmetric points,
at each point $\bm k_\parallel$ of the surface BZ,
one can always attribute an effective 1D Hamiltonian, 
which is actually nothing but $h_{3D} (k_x, \bm k_\parallel)$.
Then, at each point $\bm k_\parallel$
one solves this 1D Hamiltonian
with an open (or semi-infinite) boundary condition, and in some cases
one finds edge solutions as a mid gap state, generally at $E=\bar{E}_\pm\neq 0$, 
while in other cases one may find only bulk solutions.
Even for a given set of model parameters ($m_0$, $m_{2x}$, etc.)
this depends on the value of $\bm k_\parallel$, since
it can happen, and actually it almost always happens that an edge state 
merge and disappear into the bulk spectrum at some finite $\bm k_\parallel$.
The 
spectrum of the surface state,
typically taking a conic shape (a Dirac cone) around at least one of the four symmetric points
is found as a locus of the energy $\bar{E}_\pm (\bm k_\parallel)$
at which an edge state (if exists) appears as a solution of the corresponding 1D problem.

\begin{figure}
(a)
\includegraphics[width=75mm]{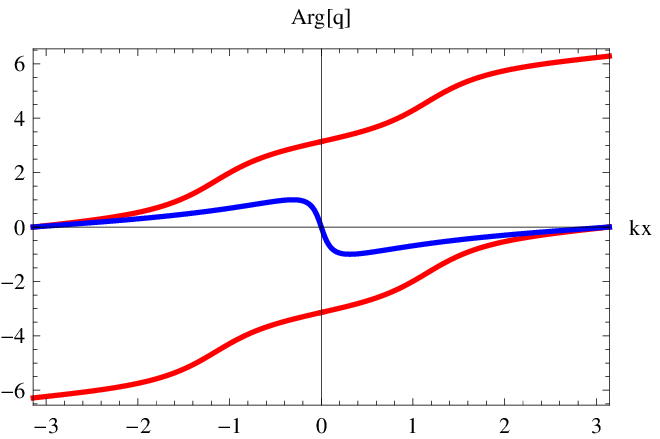}
(b)
\includegraphics[width=55mm]{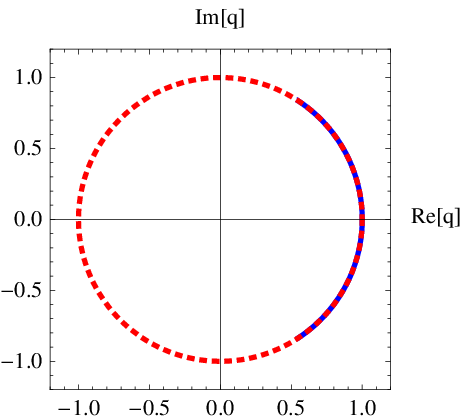}
\\
(c)
\includegraphics[width=100mm]{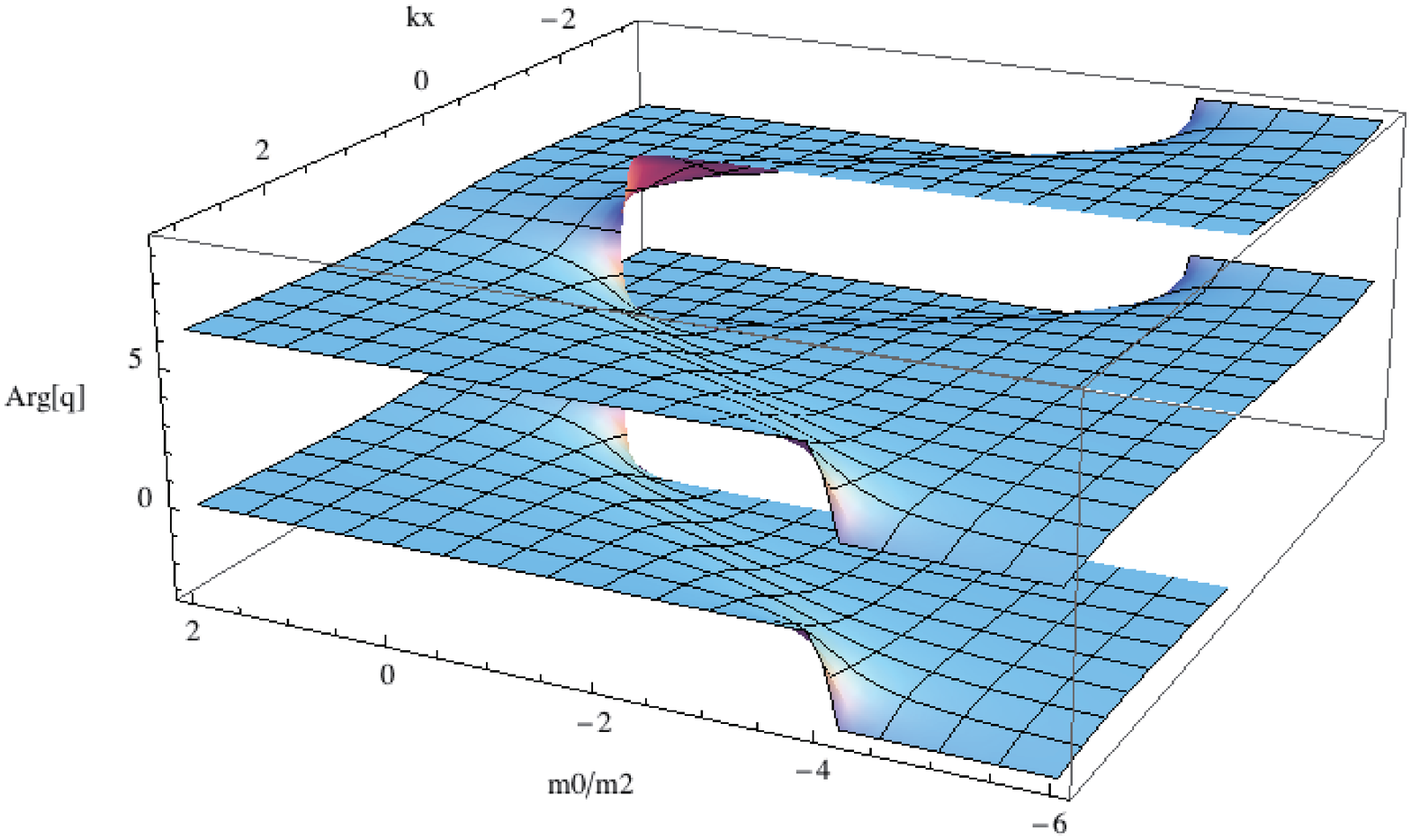}
\caption{Topological protection, or winding in the 1D model.
(a) $\arg q$ plotted as a function of $k_x$ in the trivial ($m_0/m_{2x}=0.1$, blue curve)
and nontrivial ($m_0/m_{2x}= - 1.3$, red curve) phases.
(b) locus of the points: $({\rm Re}\ q, {\rm Im}\ q)$ when $k_x$ sweeps once
the entire Brillouin zone:
$k_x \in [-\pi,\pi]$
[blue: $m_0/m_{2x}=0.1$, (red, dotted): $m_0/m_{2x}= - 1.3$].
(c) global behavior of $\arg q$ in the $(m_0/m_{2x}, k_x)$-plane.
When $m_0/m_{2x} \in [-4,0]$,
the winding number ${\cal N}_1$ as defined in Eq. (\ref{N1}) becomes nonzero.
}
\label{winding}
\end{figure}

\section{Winding properties of the 1D model}

Let us rewrite the effective 1D tight-binding Hamiltonian $h_{1D} (k_x)$
as
\begin{equation}
h_{1D} (k_x) =   m (k_x) \tau_z + t_x \sin k_x \tau_x + \epsilon (k_x) 1_2,
\label{h_1D}
\end{equation}
where
\begin{eqnarray}
m (k_x) &=& m_0 + 2 m_{2x} (1- \cos k_x),
\label{mass_1D} \\
\epsilon (k_x) &=& \epsilon_0 + 2 \epsilon_{2x} (1- \cos k_x).
\label{epsilon_1D}
\end{eqnarray}
That is to say, focusing on the case in which 
the surface Dirac cone appears at $\bm k_\parallel = \bar{\Gamma}$,
we replaced 
$\tilde{m}_0 (\bm k_\parallel)$ in Eq. (\ref{m0_tilde})
with $m_0$.
Other cases with surface Dirac cones at different symmetric points
can be discussed by reinterpreting what we we show below,
{\it i.e.}, by replacing $m_0$ with 
$\tilde{m}_0 (\bm k_\parallel)$ given in Eq. (\ref{m0_tilde2}).

To make explicit the winding properties of 1DTI Hamiltonian
(\ref{h_1D})
it is better to adopt a slightly different choice of $\gamma$-matrices
in which three elements of $\tau$ Pauli matrices are permuted as
$\tau_z\rightarrow\tau_x$ and
$\tau_x\rightarrow\tau_y$.
By setting also $\epsilon (k_x)=0$, one can rewrite $h_{1D}$ in the
following convenient form with vanishing diagonal components:
\begin{eqnarray}
\tilde{h}_{1D} (k_x) &=&  m (k_x) \tau_x + t_x \sin k_x \tau_y
\nonumber \\
&=&
\left[
\begin{array}{cc}
0 & q(k_x)\\
q(k_x)^* & 0   
\end{array}
\right],
\label{h_1D2}
\end{eqnarray}
where
$q(k_x) =m (k_x) -i t_x \sin k_x$. 
Then, by choosing an appropriate branch for $\phi = \arg q$ such that
$\phi$ becomes continuous over the entire BZ: $k_x \in [-\pi, \pi]$
one can introduce a winding number ${\cal N}_1$ such that
\begin{equation}
{\cal N}_1 = {\phi (k_x =\pi) -  \phi (k_x =-\pi)\over 2\pi},
\label{N1}
\end{equation}
associated with a mapping from $S_1$ to $S_1$ 
[mapping from the 1D BZ to a (unit) circle in the complex $q$-plane].
The nontrivial winding of $q(k_x)$ is depicted in FIG. 1.
One can indeed explicitly verify
\begin{equation}
{\cal N}_1 = \left\{
\begin{array}{ll}
1 & (-4<m_0/m_{2x}<0)\\
0 & ({\rm otherwise})\\
\end{array}
\right..
\end{equation}
As we see explicitly in the next section (see also FIG. 2, the phase diagram determined by 
the presence/absence/nature of the edge solution),
whenever the winding number ${\cal N}_1$ takes a nontrivial value (${\cal N}_1 =1$),
there appears a zero-energy mid-gap states in the spectrum, 
bound in real space to the edge of the system
(bulk-edge correspondence).
If $\epsilon (k_x) \neq 0$, the edge state still exists, but it does not appear
necessarily at the zero energy.
According to the periodic table, or the tenfold way
\cite{Schnyder_PRB,Ryu_NJP,Kitaev_AIP,Schnyder_AIP}
class DIII models
[corresponding to $\epsilon (k_x)=0$ in the present case]
in 1D
should have a topological excitation
protected by a topological number of type $\mathbb{Z}$,
while
class AII models
[corresponding to $\epsilon (k_x)\neq 0$ in the present case]
should have no protected topological excitation.
In the terminology of the tenfold way,
what is counted as a ``topological excitation''
seems to be only the bound state that appears at zero energy.
While, the distinction between the two cases here does not
seem to be so essential. 
In any case
that is how our analysis corresponds to theirs.

\begin{table}[htdp]
\caption{Band inversion in the 1D model.
$m_{2x}>0$ is assumed.
Comparison of bulk band indices
$\delta_{k_x}$ and the winding number ${\cal N}_1$.}
\begin{center}
\begin{tabular}{cccccc}
\hline\hline
&\ \ \ & $\delta_{k_x=0}$ & $\delta_{k_x=\pi}$ &\ \ \ & ${\cal N}_1$ \\
\hline
$0<m_0/m_{2x}$ && $+$ & $-$ && $0$ \\
$-4<m_0/m_{2x}<0$ && $-$ & $-$ && $1$ \\
$m_0/m_{2x}<-4$ && $-$ & $+$ && $0$\\
\hline\hline
\end{tabular}
\end{center}
\label{default}
\end{table}%

It should be also noticed that the switching of this winding number ${\cal N}_1$
is related to the band inversion at the two symmetric points:
$k_x =0$ and $k_x =\pi$.
Indeed, one can verify (see Table 1)
\begin{equation}
-{\cal N}_1 = {1\over 2}\delta_{k_x =0} +{1\over 2}\delta_{k_x =\pi},
\end{equation}
where the band indices
$\delta_{k_x}$
can be determined, for example, 
in the following way.
Re-expanding the tight-binding Hamiltonian (\ref{h_1D})
in the vicinity of the $k_x=k_x^{(0)}$
into a Dirac form,
one can express it as
\begin{equation}
h_{1D} =\hat{m}_0 \tau_z + \hat{t}_x p_x \tau_x + \hat{\epsilon}_0 1_2,
\end{equation}
in an approximation keeping only the terms up to linear order in $p_x$
($\bm k\cdot\bm p$-approximation),
where $k_x=k_x^{(0)}+p_x$.
Then one can define $\delta_{k_x}$ such that
\begin{equation}
\delta_{k_x} = {\rm sgn} (\hat{m}_0)\ 
{\rm sgn} (\hat{t}_x).
\end{equation}
In the following, we assume, without loss of generality,
$t_x>0$, and also $m_{2x}>0$,
\footnote{It is sometimes convenient to measure everything in units of $m_{2x}$.}
then $m_0>0$ (the case of a normal gap)
corresponds to a trivial phase with
$\delta_{k_x =0}=+1$ and $\delta_{k_x =\pi}=-1$ 
(see Table 1).

\section{Edge solutions in the semi-infinite geometry}

\subsection{Preliminaries}
In real space, the 1D Wilson-Dirac tight-binding Hamiltonian
given as in
Eqs. (\ref{h_1D}), (\ref{mass_1D}) and (\ref{epsilon_1D}),
can be represented by the following matrix:
\begin{equation}
H = \left[
\begin{array}{ccccccc}
\ddots & \ddots &&&& \\ 
\ddots & M & \Gamma^\dagger &&& \\ 
&\Gamma & M & \Gamma^\dagger && \\ 
&& \Gamma & M & \Gamma^\dagger &\\ 
&&& \Gamma & M &\ddots \\
&&&&\ddots & \ddots
\end{array}
\right],
\end{equation}
where $M= (\epsilon_0 + 2 \epsilon_{2x}) 1_2 + (m_0 + 2 m_{2x}) \tau_z$ represent diagonal blocks,
while 
\begin{equation}
\Gamma = -m_{2x}\tau_z + i{t_x\over 2}\tau_x  -\epsilon_{2x} 1_2,
\end{equation}
and $\Gamma^\dagger$ are (nearest-neighbor) hopping matrix elements.
The corresponding eigenvalue equation,
\begin{equation}
H\psi =E\psi,\
\psi = \left[
\begin{array}{c}
\vdots \\ \psi_0 \\ \psi_1 \\ \psi_2 \\ \psi_3 \\ \vdots  
\end{array}
\label{eigen}
\right]
\end{equation}
has a solution of the form:
\begin{equation}
\psi_j = \psi_j [\rho] = \rho^j \bm u.
\label{psi_rho}
\end{equation}
Here,
$\bm u$ is a two-component (eigen) spinor, 
satisfying 
\begin{equation}
h[\rho] \bm u = E \bm u,
\label{eigen_rho1}
\end{equation}
where
\begin{equation}
h [\rho] =  \epsilon [\rho] 1_2 + m [\rho] \tau_z + t_x {\rho -\rho^{-1}\over 2i} \tau_x,
\label{h_rho}
\end{equation}
and
$m [\rho] = m_0 + 2 m_{2x} (1-{\rho +\rho^{-1}\over 2})$,
$\epsilon [\rho] = \epsilon_0 + 2 \epsilon_{2x} (1- {\rho +\rho^{-1}\over 2})$.
Eq. (\ref{eigen_rho1}) with
Eq. (\ref{h_rho}) is nothing but the eigenvalue equation for $h(k_x)$
in momentum space,
if $\rho$ is expressed as $\rho = e^{i k_x}$.
For a given energy $E$,
the secular equation, $\det (h[\rho] - E) = 0$
gives, generally,
four solutions for $\rho = \rho_A, \rho_B, \rho_C, \rho_D$,
and the corresponding eigenvectors, $u=u_A, u_B, u_C, u_D$.

Under a periodic boundary condition:
$\psi_N=\psi_0$,
$\rho^N=1$, implying $|\rho|=1$,
or $\rho = e^{i k_x}$ with $k_x$: real and given as
$k_x = {2\pi \over N}n$ with $n=0, \pm 1, \pm 2, \cdots$.
For a given $k_x$,
the corresponding energy eigenvalue is given as
\begin{equation}
E=E(k_x)=\epsilon (k_x)\pm\sqrt{m(k_x)^2 + t^2 \sin^2 k_x}.
\end{equation}

\begin{figure}
\includegraphics[width=8cm]{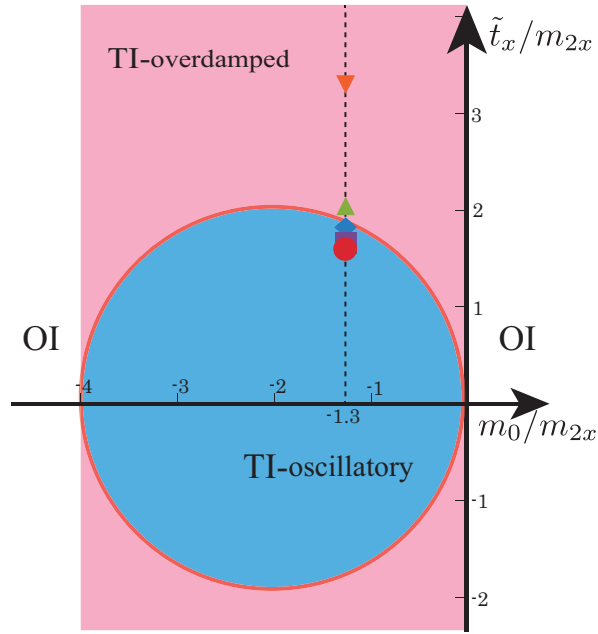}
\caption{Phase diagram of the 1D topological insulator
in the space of mass and hooping parameters
[$(m_0/m_{2x}, \tilde{t}_x/m_{2x})$-plane].
As given in Eq. (\ref{tilde_t})
$\tilde{t}_x$ is a function of $\epsilon_{2x}$.
The latter encodes 
asymmetry of the valence and conduction bands
[see Eq. (\ref{epsilon})].
}
\label{PD}
\end{figure}

\subsection{Construction of the edge solutions}
Let us consider the case of a semi-infinite geometry: $x>0$, 
in which the edge solution $\psi$ as introduced in Eq. (\ref{eigen})
has a finite amplitude $\psi_j$ on site $x=j\ge 1$
that decays also exponentially as $j\rightarrow +\infty$, {\it i.e.},
\begin{equation}
\psi_0 = 
\left[
\begin{array}{c}
0 \\ 0 
\end{array}
\right],\ 
\psi (x\rightarrow +\infty) = 
\left[
\begin{array}{c}
0 \\ 0 
\end{array}
\right],
\label{bc_semi}
\end{equation}
Such a solution is given by a difference of two solutions 
of the type given in Eq. (\ref{psi_rho}):
\begin{equation}
\psi_j= \left(\rho_A^j - \rho_B^j\right) \bm u,
\label{wf_semi}
\end{equation}
with
$\rho_A\neq\rho_B$, $|\rho_A|<1$, $|\rho_B|<1$.
Also, ${\bm u}_A ={\bm u}_B\equiv {\bm u}$ is implicit in Eq. (\ref{wf_semi});
therefore, in the following
we search for solutions of the type of  Eq. (\ref{psi_rho})
with different $\rho$ but with the same eigenspinor $\bm u$.
In other words, we need a simultaneous solution of
\begin{eqnarray}
& (h[\rho_A] -E)\ {\bm u} = 
\left[
\begin{array}{c}
0 \\ 0 
\end{array}
\right],
\nonumber \\
& (h[\rho_B] -E)\ {\bm u} =
\left[
\begin{array}{c}
0 \\ 0
\end{array}
\right].
\label{eigen_rho2}
\end{eqnarray}
These can be rewritten in the following form:
\begin{eqnarray}
&
{h[\rho_A] -E\over \gamma [\rho_A]}\ {\bm u}=
\left[\begin{array}{cc}
A_+ & -i\\ 
-i & - A_-
\end{array}
\right]
{\bm u} = 
\left[
\begin{array}{c}
0 \\ 0
\end{array}
\right],
\nonumber \\
&
{h[\rho_B] -E\over \gamma [\rho_B]}\ {\bm u}=
\left[\begin{array}{cc}
B_+ & -i\\ 
-i & - B_-
\end{array}
\right] 
{\bm u} =
\left[
\begin{array}{c}
0 \\ 0
\end{array}
\right],
\label{eigen_AB}
\end{eqnarray}
where
\begin{eqnarray}
A_+ = {m_A - (E - \epsilon_A) \over\gamma_A},\
B_+ = {m_B - (E - \epsilon_B) \over\gamma_B}
\nonumber \\
A_- = {m_A + E- \epsilon_A \over\gamma_A},\
B_- = {m_B + E - \epsilon_B \over\gamma_B}.
\end{eqnarray}
Here, 
$m_A$, $m_B$, $\epsilon_A$ and $\epsilon_B$,
are short-hand notations for $m[\rho_A]$, $m[\rho_B]$, $\epsilon [\rho_A]$ and $\epsilon [\rho_B]$.
We have also introduced,
\begin{equation}
\gamma [\rho] = t_x {\rho - \rho^{-1} \over 2},
\end{equation}
and
$\gamma_A =\gamma [\rho_A]$, $\gamma_B =\gamma [\rho_B]$.
Clearly, the two eigenvalue equations
share the same eigenvector $\bm u$
only when $A_+=B_+$ and $A_-=B_-$, implying
\begin{equation}
{m_A - (E - \epsilon_A)\over m_A +E - \epsilon_A} = 
{m_B - (E - \epsilon_B)\over m_B + E - \epsilon_B}.
\label{cond_en1}
\end{equation}
This condition imposed by the semi-infinite geometry, or by the boundary 
condition (\ref{bc_semi})
plays the role of determining the energy $E$ at which an edge state appears;
Eq. (\ref{cond_en1}) first simplifies to
\begin{equation}
m_A (E-\epsilon_B) = m_B (E-\epsilon_A).
\label{cond_en2}
\end{equation}
Then, recalling
$m_A = m_0 + m_{2x} k_A^2$, $m_B = m_0 + m_{2x} k_B^2$, 
$\epsilon_A = \epsilon_0 + \epsilon_{2x} k_A^2$ and $\epsilon_B = \epsilon_0 + \epsilon_{2x} k_B^2$, 
where $k_A^2$ and $k_B^2$ are
short-hand notations for
$k_A^2=2(1-{\rho_A +\rho_A^{-1}\over 2})$ and
$k_B^2=2(1-{\rho_B +\rho_B^{-1}\over 2})$,
one finds,
\begin{equation}
(k_A^2 -k_B^2)\{m_0 \epsilon_{2x} + m_{2x} (E - \epsilon_0)\}=0.
\end{equation}
Since $\rho_A \neq\rho_B$, this means,\cite{Shen_NJP}
\begin{equation}
E = \epsilon_0 - {m_0 \over m_{2x}}\epsilon_{2x}
\equiv \bar{E},
\label{E0}
\end{equation}
and this is the energy at which an edge state appears.
In the special case of $\epsilon (k_x) =0$ (particle-hole symmetric case),
this reduces to a simple zero-energy condition: $E=0$,
while here, in the presence of $\epsilon (k_x)\neq 0$,
this condition is relaxed.
Eq. (\ref{E0}) also implies that at this energy $E=\bar{E}$
the two independent parameters
$\epsilon [\rho] - E$ and $m[\rho]$ have the same functional form:
\begin{equation}
\epsilon [\rho] - \bar{E} = {\epsilon_{2x}\over m_{2x}} m[\rho].
\label{c-E}
\end{equation}

Let us go back to the eigenvalue equation:
Eq. (\ref{eigen_rho1}), or Eqs. (\ref{eigen_rho2}), 
and find the corresponding eigenvector.
Let us recall
\begin{equation}
h[\rho]-E=
\left[\begin{array}{cc}
m[\rho]+\epsilon [\rho]-E & -i\gamma [\rho]\\ 
-i\gamma [\rho] & - (m[\rho]-\epsilon [\rho]+E)
\end{array}
\right]
\end{equation}
Then, from the secular equation: $\det (h[\rho] - E) = 0$, one must have,
\begin{equation}
\gamma [\rho] = \pm\sqrt{m[\rho]^2 -(E-\epsilon [\rho])^2}
\equiv \gamma_{1,2},
\label{h-E}
\end{equation}
The subscripts of $\gamma$ corresponds to the two choices of sign
in front of the square root in the middle expression.
At $E=\bar{E}$,
Eq. (\ref{c-E}) implies
\begin{equation}
\gamma [\rho] =
\gamma_{1,2} [\rho] = \pm\sqrt{1-{\epsilon_{2x}^2\over m_{2x}^2}}m[\rho],
\label{gamma}
\end{equation}
{\it i.e.}, $\gamma [\rho]$ also shares the same functional form as $m[\rho]$.
But here, let us continue expressing $\gamma$'s as in Eq. (\ref{h-E}),
and rewrite Eqs. (\ref{eigen_AB}) as
\begin{equation}
{h_{1,2} -E \over\gamma_{1,2}}\ {\bm u}_{1,2}=
\left[\begin{array}{cc}
\pm\sqrt{m+\epsilon-E \over m-\epsilon+E} & -i \\ 
-i & \mp\sqrt{m-\epsilon+E\over m+\epsilon-E} 
\end{array}
\right]
\ {\bm u}_{1,2}
= 
\left[
\begin{array}{c}
0 \\ 0
\end{array}
\right].
\end{equation}
Here, we introduced the notation $h_{1,2}$ such that
$h=h_{1,2}$ when $\gamma =\gamma_{1,2}$.
The corresponding eigenvectors ${\bm u}_{1,2}$ are found
(up to normalization, {\it i.e.}, $|{\bm u}_1|^2=|{\bm u}_2|^2=1$) as
\begin{eqnarray}
\bm u_1 &=& 
{1\over\sqrt{2m}}
\left[
\begin{array}{c}
\sqrt{m-\epsilon+E}\\
-i\sqrt{m+\epsilon-E}
\end{array}
\right],
\label{u1} \\
\bm u_2 &=& 
{1\over\sqrt{2m}}
\left[
\begin{array}{c}
\sqrt{m-\epsilon+E}\\
i\sqrt{m+\epsilon-E}
\end{array}
\right].
\label{u2}
\end{eqnarray}
where $m=m[\rho]$ and $\epsilon =\epsilon [\rho]$
are functions of $\rho$, and through $\rho=\rho (E)$
they are also functions of $E$.
At $E=\bar{E}$ these reduce to
\begin{eqnarray}
\bm u_1^{(0)} &=& 
{1\over\sqrt{2}}
\left[
\begin{array}{c}
\sqrt{1-{\epsilon_{2x}\over m_{2x}}}\\
-i\sqrt{1+{\epsilon_{2x}\over m_{2x}}}
\end{array}
\right],
\label{u10} \\
\bm u_2^{(0)} &=& 
{1\over\sqrt{2}}
\left[
\begin{array}{c}
\sqrt{1-{\epsilon_{2x}\over m_{2x}}}\\
i\sqrt{1+{\epsilon_{2x}\over m_{2x}}}
\end{array}
\right].
\label{u20}
\end{eqnarray}

The identity:
$\gamma [\rho] = \gamma_1 [\rho] =
\sqrt{1-{\epsilon_{2x}^2\over m_{2x}^2}}m[\rho]$
in Eq. (\ref{gamma}),
can be viewed
as a quadratic equation for $\rho$,
{\it i.e.},
$a\rho^2 + b\rho + c=0$
with
\begin{equation}
a=m_{2x} +\tilde{t}_x/2,\
b=-(m_0 +2m_{2x}),\
c=m_{2x} -\tilde{t}_x/2,
\label{pqr}
\end{equation}
where
\begin{equation}
\tilde{t}_x={t_x\over\sqrt{1-{\epsilon_{2x}^2\over m_{2x}^2}}}
\label{tilde_t}
\end{equation}
represents rescaled hopping.
Let us name the two solutions of this quadratic equation as
\begin{equation}
\rho=\rho_{1\pm}=
{m_0 +2 m_{2x} \pm \sqrt{(m_0 +2 m_{2x})^2 -4(m_{2x}^2 - \tilde{t}_x^2/4)}\over 2(m_{2x} + \tilde{t}_x/2)}.
\label{rho1}
\end{equation}

The second choice for $\gamma$ in Eq. (\ref{gamma}), {\it i.e.},
$\gamma [\rho] = \gamma_2 [\rho] =
-\sqrt{1-{\epsilon_{2x}^2\over m_{2x}^2}}m[\rho]$
replaces Eqs. (\ref{pqr}) with
\begin{equation}
a=m_{2x} -\tilde{t}_x/2,\
b=-(m_0 +2m_{2x}),\
c=m_{2x} +\tilde{t}_x/2,
\end{equation}
{\it i.e.}, $\tilde{t}_x$ is replaced with $-\tilde{t}_x$.
This leads to the second set of solutions for $\rho$:
\begin{equation}
\rho=\rho_{2\pm}=
{m_0 +2 m_{2x} \pm \sqrt{(m_0 +2 m_{2x})^2 -4(m_{2x}^2 - \tilde{t}_x^2/4)}\over 2(m_{2x} - \tilde{t}_x/2)}.
\label{rho2}
\end{equation}
The case of $m_{2x}\pm\tilde{t}_x/2=0$ may need a separate consideration.

Case of $D=b^2-4ac <0$: a pair of complex conjugate solutions;
the two solutions for $\rho$
become a pair of mutually conjugate complex numbers,
\begin{equation}
\rho={-b \pm i \sqrt{4ac - b^2} \over 2a}.
\label{complex}
\end{equation}
Therefore,
\begin{equation}
|\rho|^2={b^2 + 4ac - b^2 \over 4 a^2}={c\over a}
={m_{2x}\mp\tilde{t}_x/2\over m_{2x}\pm\tilde{t}_x/2}
\end{equation}
If one sets the parameters such that $m_{2x}>0$ and $t_x>0$, 
one must have
\begin{equation}
|\rho_{1\pm}|<1,\   |\rho_{2\pm}|>1.
\label{ineq}
\end{equation}
Then, since
$\rho =\rho_{1\pm}$
share the same eigenspinor $\bm u_1^{(0)}$,
choosing in Eq. (\ref{wf_semi})
as $\rho_A =\rho_{1+}$ and $\rho_B =\rho_{1-}$,
we find that
\begin{equation}
\psi_{\rm semi} = \left[
\begin{array}{c}
\psi_0 \\ \psi_1 \\ \psi_2 \\ \psi_3 \\ \vdots  
\end{array}
\right]
\label{psi_semi}
\end{equation}
with 
\begin{eqnarray}
\psi_j &=&{\cal N} \left(\rho_{1+}^j - \rho_{1-}^j\right) \bm u_1
\nonumber \\
&=& {{\cal N}\over \sqrt{2}}\
(\rho_{1+}^j - \rho_{1-}^j)
\left[
\begin{array}{c}
\sqrt{1-{\epsilon_{2x}\over m_{2x}}}\\ 
-i\sqrt{1+{\epsilon_{2x}\over m_{2x}}}
\end{array}
\right].
\label{sol_semi}
\end{eqnarray}
(for $j=0,1,2,3,\cdots$)
is the edge/surface solution in the semi-infinite geometry that appears at
$E=\bar{E}=\epsilon_0 + {m_0 \over m_{2x}}\epsilon_{2x}$ 
as given in Eq. (\ref{E0}).
${\cal N}$ is a normalization constant.
A pair of complex conjugate solutions for $\rho$ implies that the profile of
the edge/surface wave function as specified by Eqs. (\ref{psi_semi}) and (\ref{sol_semi})
shows a {\it damped oscillation}. 
FIG. \ref{PD} depicts the phase diagram of our 1D topological insulator, which indicates
the region of parameters in which such an edge/surface wave function with
a damped oscillatory profile appears.
As implied in
Eqs. (\ref{rho1}) and (\ref{rho2})
such a region is represented by the interior of
a circle:
\begin{equation}
\left({m_0\over m_{2x}}+2\right)^2+
\left({\tilde{t}_x \over m_{2x}}\right)^2 = 4,
\label{circle}
\end{equation}
in the $(m_0/m_{2x}, \tilde{t}_x/m_{2x})$-space.
In FIG. \ref{PD}
this region is denoted as ``TI-oscillatory'', and
painted in pale blue.

Case of $D=b^2-4ac >0$: two real solutions.
In this case the two solutions for $\rho$ in Eqs. (\ref{rho1}) and (\ref{rho2})
become two real solutions.
One can verify that the inequality (\ref{ineq}) still holds.
The edge solution: (\ref{psi_semi}) with (\ref{sol_semi}) is also formally unchanged, but now
the wave function is {\it overdamped}.
The corresponding parameter region 
in the $(m_0/m_{2x}, \tilde{t}_x/m_{2x})$-space
is the intersection of the exterior of the circle represented by Eq. \ref{circle}
and the strip region: $-4\le m_0/m_{2x}\le 0$.
In the phase diagram of FIG. \ref{PD}
this region is denoted as ``TI-overdamped'', and painted in pink.

\begin{figure}
\includegraphics[width=8cm]{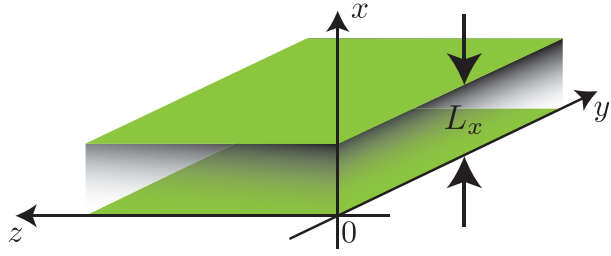}
\caption{The slab geometry.
TI nanofilm of a thickness $L_x$ [$=$ the number of (quintuple) layers stacked].}
\label{slab}
\end{figure}

\section{Finite-size energy gap in the slab}

Let us now generalize
the previous construction of the surface wave function
developed in the case of semi-infinite geometry
to a more realistic case of the slab 
with a finite thickness $L_x$;
the topological insulator
occupies the region of $1\le x\le L_x$
(see FIG. \ref{slab}).
The same applies to our 1D effective model. 
As a result of an explicit construction, we find
an (unusually) intimate relation 
between the magnitude of the gap and the profile of the surface wave function.
\cite{weak_eo}
Here, instead of Eqs. (\ref{bc_semi})
we apply the following boundary condition:
\begin{equation}
\psi_0= \left[
\begin{array}{c}
0 \\ 0
\end{array}
\right],\ 
\psi_{L_x+1} = \left[
\begin{array}{c}
0 \\ 0
\end{array}
\right].
\label{bc_slab}
\end{equation}
In this slab geometry, 
the asymptotic condition at $x\rightarrow +\infty$
[the second condition in Eqs. (\ref{bc_semi})]
is no longer effective;
so the solution of the boundary problem is
expressed as a linear combination of the
exponentially increasing and decreasing solutions.

We have seen earlier that 
$\bar{E}$ given as in Eq. (\ref{E0}) is the energy of the edge solution 
in the semi-infinite geometry.
At this energy $E=\bar{E}$, the edge solution: 
Eq. (\ref{psi_semi}) with Eq. (\ref{sol_semi})
was constructed by superposing two exponentially decreasing functions
specified by $\rho_{1\pm}$ [given in Eq. (\ref{rho1})]
and by ${\bm u}_1^{(0)}$ [given in Eqs. (\ref{u10})].
The remaining set of solutions 
[that are not compatible with the boundary condition (\ref{bc_semi})
and specified by
$\rho_{2\pm}$ as given in Eq. (\ref{rho2})
and by ${\bm u}_2^{(0)}$ as given in Eqs. (\ref{u20})]
are exponentially increasing functions.
At $E=\bar{E}$, 
these two purely exponentially decreasing and increasing
solutions were orthogonal to each other
[see Eqs. (\ref{u10}) and (\ref{u20})]. 
This, in turn, implies that in the slab geometry
the energy $E$ of the edge solution must deviate from $\bar{E}$;
in the present case of a slab, 
on needs,
to cope with the boundary condition (\ref{bc_slab}),
to superpose both the exponentially decreasing and increasing
solutions, but at $E=\bar{E}$ this was precisely not possible 
because of the orthogonality of
${\bm u}_1^{(0)}$ and ${\bm u}_2^{(0)}$.
In the following,
we attempt to construct an edge solution compatible
with the boundary condition (\ref{bc_slab})
by superposing exponentially decreasing and increasing solutions
at $E$ away from but still close to $\bar{E}$.

To be more specific, we first expand the eigenvector ${\bm u}_{1,2}$ 
as given in Eqs. (\ref{u1}) and (\ref{u2}) around $E=\bar{E}$, then by taking into account
the first order corrections to ${\bm u}_{1,2}^{(0)}$ (in powers of $E-\bar{E}$)
we attempt to construct an edge solution compatible
with the boundary condition (\ref{bc_slab}).
Denoting as $E=\bar{E} + \delta E$, 
let us also expand $m=m[\rho]$ and $\epsilon =\epsilon[\rho]$ as
$m=\bar{m}+\delta m$ and $\epsilon =\bar{\epsilon}+\delta\epsilon$.
Here, $\delta m$ and $\delta\epsilon$ denote corrections of order $\delta E$.
Taking these into account one can expand
each component of Eqs. (\ref{u1}) and (\ref{u2}) as
\begin{eqnarray}
\sqrt{m-\epsilon+E\over m}
&\simeq&
\sqrt{\left(1-{\epsilon_{2x}\over m_{2x}}\right)}
\left[1+{\delta E - \delta\epsilon+{\epsilon_{2x}\over m_{2x}}\delta m
\over 2\bar{m}\left(1-{\epsilon_{2x}\over m_{2x}}\right)} \right]
\nonumber \\
&=&
\sqrt{\left(1-{\epsilon_{2x}\over m_{2x}}\right)}
\left[1+{\delta E\over 2\bar{m}\left(1-{\epsilon_{2x}\over m_{2x}}\right)} \right],
\label{expand1}
\end{eqnarray}
\begin{eqnarray}
\sqrt{m+\epsilon-E\over m}
&\simeq&
\sqrt{\left(1+{\epsilon_{2x}\over m_{2x}}\right)}
\left[1-{\delta E - \delta\epsilon+{\epsilon_{2x}\over m_{2x}}\delta m
\over 2\bar{m}\left(1+{\epsilon_{2x}\over m_{2x}}\right)} \right]
\nonumber \\
&=&
\sqrt{\left(1+{\epsilon_{2x}\over m_{2x}}\right)}
\left[1-{\delta E\over 2m\left(1+{\epsilon_{2x}\over m_{2x}}\right)} \right].
\label{expand2}
\end{eqnarray}
In deriving Eqs. (\ref{expand1}), (\ref{expand2})
we have used the relation (\ref{c-E}), and in simplying the expressions,
we have also noticed,
\begin{equation}
\delta\epsilon ={\epsilon_{2x}\over m_{2x}}\delta m.
\label{delta_m}
\end{equation}
This follows immediately from the definition of $m$ and $\epsilon$:
$m[\rho]=m_0+2m_{2x}(\rho+\rho^{-1})$,
$\epsilon[\rho]=\epsilon_0+2\epsilon_{2x}(\rho+\rho^{-1})$,
since these imply,
$\delta m[\rho]=2m_{2x}\delta (\rho+\rho^{-1})$, and
$\epsilon[\rho]=2\epsilon_{2x}\delta (\rho+\rho^{-1})$,
suggesting Eq. (\ref{delta_m}).
Here, $\rho$ should be also expanded up to linear order in $\delta E$ as
$\rho=\bar{\rho}+\delta\rho$.
Using Eqs. (\ref{expand1}), (\ref{expand2}),
one can rewrite ${\bm u}_{1,2}$ as
\begin{eqnarray}
\bm u_1 &=&
{1\over\sqrt{2m}}
\left[
\begin{array}{c}
\sqrt{m-\epsilon+E}\\
-i\sqrt{m+\epsilon-E}
\end{array}
\right]
\nonumber \\
&\simeq&
{1\over\sqrt{2}}
\left[
\begin{array}{c}
\sqrt{1-{\epsilon_{2x}\over m_{2x}}}\\
-i\sqrt{1+{\epsilon_{2x}\over m_{2x}}}
\end{array}
\right]
+{\delta E\over 2\bar{m}[\rho] \sqrt{1-\left({\epsilon_{2x}\over m_{2x}}\right)^2}}
{1\over\sqrt{2}}
\left[
\begin{array}{c}
\sqrt{1+{\epsilon_{2x}\over m_{2x}}}\\
i\sqrt{1-{\epsilon_{2x}\over m_{2x}}}
\end{array}
\right]
\nonumber \\
&\equiv&
{\bm u}_1^{(0)}
+{\delta E\over 2\bar{m}[\rho] \sqrt{1-\left({\epsilon_{2x}\over m_{2x}}\right)^2}}
\tilde{\bm u}_2,
\label{u1_E}
\end{eqnarray}
\begin{eqnarray}
\bm u_2 &=&
{1\over\sqrt{2m}}
\left[
\begin{array}{c}
\sqrt{m-\epsilon+E}\\
i\sqrt{m+\epsilon-E}
\end{array}
\right]
\nonumber \\
&\simeq&
{1\over\sqrt{2}}
\left[
\begin{array}{c}
\sqrt{1-{\epsilon_{2x}\over m_{2x}}}\\
i\sqrt{1+{\epsilon_{2x}\over m_{2x}}}
\end{array}
\right]
+{\delta E\over 2\bar{m}[\rho] \sqrt{1-\left({\epsilon_{2x}\over m_{2x}}\right)^2}}
{1\over\sqrt{2}}
\left[
\begin{array}{c}
\sqrt{1+{\epsilon_{2x}\over m_{2x}}}\\
-i\sqrt{1-{\epsilon_{2x}\over m_{2x}}}
\end{array}
\right]
\nonumber \\
&\equiv&
{\bm u}_2^{(0)}
+{\delta E\over 2\bar{m}[\rho] \sqrt{1-\left({\epsilon_{2x}\over m_{2x}}\right)^2}}
\tilde{\bm u}_1.
\label{u2_E}
\end{eqnarray}
The last lines define $\tilde{\bm u}_1$ and $\tilde{\bm u}_2$. 
They reduce, respectively, to ${\bm u}_1^{(0)}$ and ${\bm u}_2^{(0)}$
in the limit: $\epsilon_{2x}\rightarrow 0$, while generally this is not the case.
In Eqs. (\ref{u1_E}) and (\ref{u2_E}), 
$m$ and $\bar{m}$ are functions of $\rho$;
they take different values in the two expressions, while
even in the same expression for $\bm u_1$
they differ for $\rho =\rho_{1+}$ and for $\rho =\rho_{1-}$. 
To specify this point, we add subscripts $1\pm, 2\pm$ to $\bar{m}$ as $\bar{m}_{1\pm}$ and $\bar{m}_{2\pm}$,
and also to ${\bm u}_{1,2}$ as ${\bm u}_{1\pm}$ and ${\bm u}_{2\pm}$
in the following expressions.
Also, $\rho$ in $\bar{m}[\rho]$
in the denominator
of the second term of
Eqs. (\ref{u1_E}) and (\ref{u2_E}) is generally
a function of $E$.
But here, since the term itself is already first order in $\delta E$,
we can safely replace this with
the value of Eqs. (\ref{rho1}) and  (\ref{rho2}) found at $E=\bar{E}$.

Armed with these basic solutions, 
we can now construct a wave function 
compatible with the boundary condition (\ref{bc_slab})
in the spirit of Eq.  (\ref{sol_semi}) .
But here,
both the exponentially increasing and decreasing solutions
must be taken into account.
Let us consider the superposition:
\begin{eqnarray}
\psi_j = c_{1+} \rho_{1+}^j {\bm u}_{1+} - c_{1-} \rho_{1-}^j {\bm u}_{1-} +
c_{2+} \rho_{2+}^{j-L_x-1}{\bm u}_{2+} - c_{2-} \rho_{2-}^{j-L_x-1} {\bm u}_{2-}.
\label{psi_rho_v0}
\end{eqnarray}
Here, we have taken into account that the eigenspinor ${\bm u}_{1\pm}$
are no longer generally identical.
The coefficients $c_{1\pm}$, $c_{2\pm}$
are to be determined so that the above $\psi$
is compatible with the boundary condition (\ref{bc_slab}).
Using the expansions (\ref{u1_E}) and (\ref{u2_E}), 
we can rewrite Eq. (\ref{psi_rho_v0}) as
\begin{eqnarray}
\psi_j &=& 
\left( c_{1+} \rho_{1+}^j - c_{1-} \rho_{1-}^j \right)
{\bm u}_1^{(0)}
+
\left( c_{1+} {\rho_{1+}^j \over 2\bar{m}_{1+}} - c_{1-} {\rho_{1-}^j\over 2 \bar{m}_{1-}} \right)
{\delta E\over\sqrt{1-\left({\epsilon_{2x}\over m_{2x}}\right)^2}}
\tilde{\bm u}_2
\nonumber \\
&+&
\left(c_{2+}\rho_{2+}^{j-L_x-1} - c_{2-}\rho_{2-}^{j-L_x-1} \right)
{\bm u}_2^{(0)}
+
\left( c_{2+} {\rho_{2+}^{j-L_x-1} \over 2\bar{m}_{2+}} - c_{2-} {\rho_{2-}^{j-L_x-1} \over 2\bar{m}_{2-}} \right)
{\delta E\over\sqrt{1-\left({\epsilon_{2x}\over m_{2x}}\right)^2}}
\tilde{\bm u}_1.
\label{psi_rho_v1}
\end{eqnarray}
Let us impose the boundary condition (\ref{bc_slab})
to the above general formal from for $\psi_j$ given as in (\ref{psi_rho_v0}).
Noticing the relations
\begin{eqnarray}
{\bm u}_1^{(0)\dagger} {\bm u}_1^{(0)} &=&1,
\nonumber \\
{\bm u}_1^{(0)\dagger} \tilde{\bm u}_2^{(0)} &=& 0,
\nonumber \\
{\bm u}_1^{(0)\dagger} {\bm u}_2^{(0)} &=& - {\epsilon_{2x} \over m_{2x}},
\nonumber \\
\tilde{\bm u}_2^{(0)\dagger} {\bm u}_1^{(0)} &=&0,
\nonumber \\
\tilde{\bm u}_2^{(0)\dagger} \tilde{\bm u}_2^{(0)} &=&1,
\nonumber \\
\tilde{\bm u}_2^{(0)\dagger} {\bm u}_2^{(0)} &=& \sqrt{1-\left(\epsilon_{2x} \over m_{2x}\right)^2},
\end{eqnarray}
and at the leading order of $\rho_{1\pm}^{L_x}$, $\rho_{2\pm}^{-L_x}$ and $\delta E$
(here, we consider the case of $|\rho_{1\pm}|<1$),
the original boundary problem (\ref{bc_slab})
reduces to the following linear system for the four unknown coefficients $c_{1\pm}$ and $c_{2\pm}$:
\begin{equation}
\left[
\begin{array}{cccc}
1 & -1 & - {\epsilon_{2x} \over m_{2x}}\rho_{2+}^{-L_x-1} & {\epsilon_{2x} \over m_{2x}}\rho_{2-}^{-L_x-1}
\\
{1\over 2\bar{m}_{1+}}{\delta E\over 1-\left({\epsilon_{2x}\over m_{2x}}\right)^2}
& - {1\over 2\bar{m}_{1-}}{\delta E \over 1-\left({\epsilon_{2x}\over m_{2x}}\right)^2}
&\rho_{2+}^{-L_x-1} & - \rho_{2-}^{-L_x-1}
\\
-{\epsilon_{2x} \over m_{2x}}\rho_{1+}^{L_x+1} & {\epsilon_{2x} \over m_{2x}}\rho_{1-}^{L_x+1}
&1 & -1 
\\
\rho_{1+}^{L_x+1} & - \rho_{1-}^{L_x+1} &
{1\over 2\bar{m}_{2+}}{\delta E\over 1-\left({\epsilon_{2x}\over m_{2x}}\right)^2}
& - {1\over 2\bar{m}_{2-}}{\delta E \over 1-\left({\epsilon_{2x}\over m_{2x}}\right)^2}
\end{array}
\right]
\left[
\begin{array}{c}
c_{1+}\\ c_{1-}\\c_{2+}\\c_{2-}
\end{array}
\right]
=
\left[
\begin{array}{c}
0\\ 0\\0\\0
\end{array}
\right].
\label{secular}
\end{equation}
Finally, we solve the secular equation 
[(the determinant of the coefficient matrix of Eq. (\ref{secular})) $= 0$]
for $\delta E$
to find the magnitude the finite-size gap.
Solving the secular equation, we recall the relations such as
\begin{eqnarray}
\rho_{2\pm}^{-1} &=& \rho_{1\mp},
\nonumber \\
\bar{m}_{1\pm} =\gamma_{1\pm} &=& {\tilde{t}_x\over 2}(\rho_{1\pm} - \rho_{1\pm}^{-1})
\nonumber \\
&=& {\tilde{t}_x\over 2}(\rho_{2\mp}^{-1} - \rho_{2\mp}) = - \gamma_{2\mp} = \bar{m}_{2\mp}.
\end{eqnarray}
One finds
\begin{eqnarray}
\delta E &=&\pm
2\left\{1-\left({\epsilon_{2x}\over m_{2x}}\right)^2\right\}
{
\rho_{1-}^{L_x+1} - \rho_{1+}^{L_x+1} 
\over
{1\over \bar{m}_{1-}} - {1\over \bar{m}_{1+}}
\mp {\epsilon_{2x} \over m_{2x}}
\left(
{\rho_{1+}^{L_x +1} \over \bar{m}_{1-}}-{\rho_{1-}^{L_x+1} \over \bar{m}_{1+}} 
\right)
}
\nonumber \\
&=&\pm
2\left\{1-\left({\epsilon_{2x}\over m_{2x}}\right)^2\right\}
{
\rho_{2+}^{-L_x-1} - \rho_{2-}^{-L_x-1}
\over
{1\over \bar{m}_{2+}} - {1\over \bar{m}_{2-}}
\mp {\epsilon_{2x} \over m_{2x}}
\left(
{\rho_{2-}^{-L_x -1} \over \bar{m}_{2+}}-{\rho_{2+}^{-L_x-1} \over \bar{m}_{2-}}
\right)
}
\equiv \delta E_\pm .
\label{deltaE2}
\end{eqnarray}
Note that the second term in the denominator 
of Eq. (\ref{deltaE2})
is much smaller than the first term, 
and in most cases irrelevant.
When this is the case,
one may convince oneself by
comparing Eq. (\ref{deltaE2}) and Eq. (\ref{sol_semi}),
that the magnitude of the finite-size energy gap
$|\delta E_+ - \delta E_-| \equiv 2 E_0$
in the slab of a thickness $L_x$
is directly proportional to the amplitude of the wave function at the depth of $x=L_x+1$,
\cite{weak_eo}
{\it i.e.},
\begin{equation}
2 E_0 (L_x) \simeq
{4\over {\cal N}}
{1-\left({\epsilon_{2x}\over m_{2x}}\right)^2
\over
\left|
{1\over \bar{m}_{1+}} - {1\over \bar{m}_{1-}}
\right|
}
|\psi_{\rm semi} (x=L_x+1)|.
\label{gap_wf}
\end{equation}
Here, $\psi_{\rm semi}$ represents the surface wave function in the
semi-infinite geometry as given in Eq. (\ref{psi_semi}).
Eqs. (\ref{deltaE2}) and (\ref{gap_wf}),
together with
the formulas for the phase boundary [Eqs. (\ref{tilde_t}) and (\ref{circle})]
constitute the central result of this paper.

\begin{figure}
\includegraphics[width=160mm]{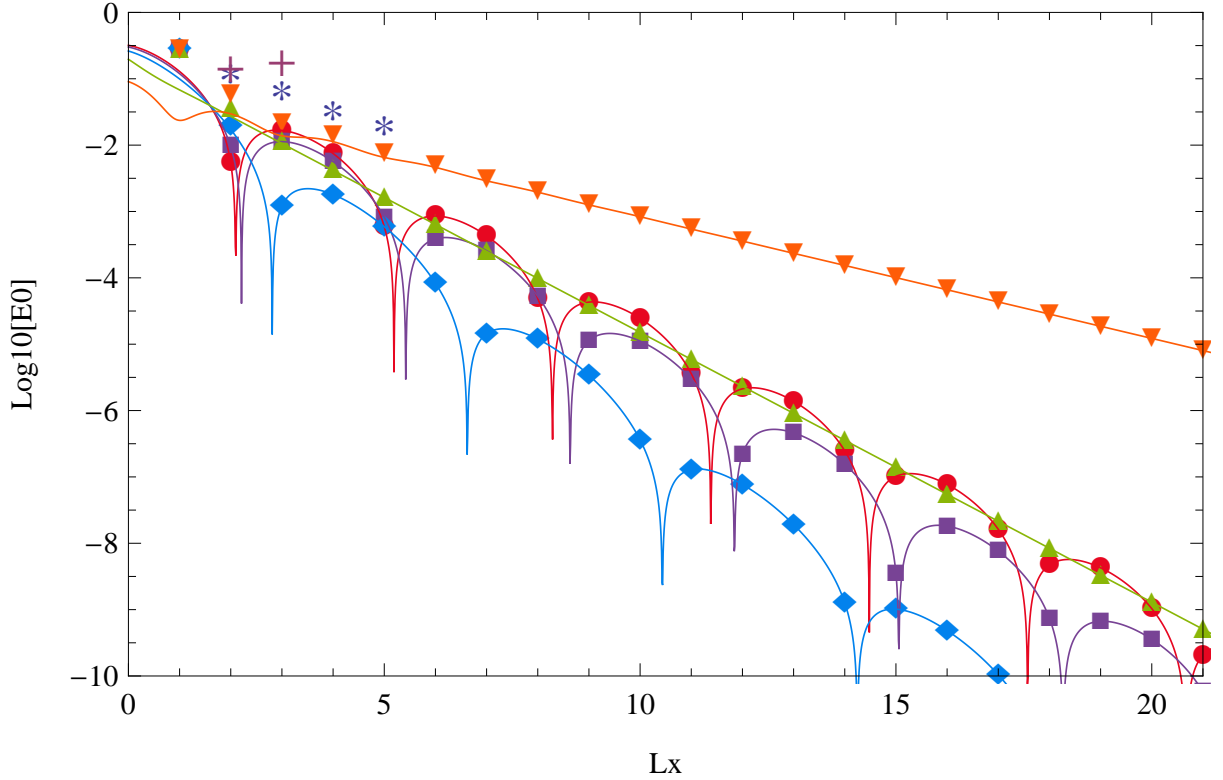}
\caption{Thickness ($L_x$) dependence of the (half) size energy gap $E_0$ in TI thin films.
Comparison of $E_0 (L_x)$ found by numerical diagonalization of the tight-binding Hamiltonian
with open boundary conditions at $x=1$ and $x=L_x$ 
[for a given set of model parameters this appears as a series of points with the same symbol and color], 
and the same dependence predicted by Eqs. (\ref{deltaE2}) 
[shown as a continuous line in the same color].
For the series of filled red points
the model parameters are chosen as given in Eqs. (\ref{params}).
In FIG. \ref{PD}
this corresponds to a point $(m_0/m_{2x}, \tilde{t}_x/m_{2x})=(-1.3, 1.51)$
as indicated by the same symbol and color.
Other set of points correspond to
$\epsilon_{2x}/m_{2x} = 0.3, 0.5, 0.7, 0.9$,
{\it i.e.}, asymmetry in the spectrum of valence and conduction bands is enhanced
from the original Bi$_2$Se$_3$ value.
These set of points are all on the (dotted) line $(m_0/m_{2x} = -1.3$ in FIG. \ref{PD},
while the value of effective hopping $\tilde{t}_x$ varies as
$\tilde{t}_x/m_{2x} = 1.57, 1.73, 2.10, 3.44$.
The corresponding points are indicated by the same symbol and color in
FIG. \ref{PD}.
Experimental values that have appeared in Refs. \cite{Hirahara}, \cite{exp_film1} 
are also shown in the figure in plus and asterisk symbols for comparison.}
\label{gap_log}
\end{figure}

\section{Comparison of analytic vs. numerical results}

To check the validity of the analyses in the preceding  sections
let us compare the finite-size energy gap obtained numerically in slab-shaped samples
with the formulas we found so far 
[Eqs. (\ref{tilde_t}), (\ref{circle}), (\ref{deltaE2}) and (\ref{gap_wf})].
Let us consider the case of following model parameters deduced from material parameters 
of Bi$_2$Se$_3$: \cite{ebi-chan}
\begin{eqnarray}
m_0&=& -0.28,\ m_{2x}=0.216,
\nonumber \\
\epsilon_0 &=& - 0.0083, \epsilon_{2x} = 0.024, 
\nonumber \\
t_x&=& 0.32,
\label{params}
\end{eqnarray}
in units of eV.
In the original 3D bulk Hamiltonian, the remaining set of parameters:
\begin{eqnarray}
& m_{2y}=m_{2z}=2.6,
\nonumber \\
&\epsilon_{2y} = \epsilon_{2z} = 1.77,
\nonumber \\
&t_y=t_z =0.8
\end{eqnarray}
are also relevant.
The set of parameters specified by
Eqs. (\ref{params})
correspond
in the phase diagram of FIG. \ref{PD}
to a point $(m_0/m_{2x}, \tilde{t}_x/m_{2x})=(-1.3, 1.51)$,
denoted in the figure by a filled red circle,
which
falls on the ``TI-oscillatory'' phase,
exhibiting a surface state with
a damped oscillatory wave function.
Note that the phase boundary between
the TI-oscillatory and TI-overdamped phases
is a circle represented by Eq. (\ref{circle})
in the $(m_0/m_{2x}, \tilde{t}_x/m_{2x})$-plane,
while
the magnitude of the effective hopping
$\tilde{t}_x$ given in Eq. (\ref{tilde_t})
is a function of $\epsilon_{2x}$.
Note that $\epsilon_{2x}$
encodes asymmetry of the spectrum
in the valence and conduction bands.
This signifies that
one can drive the system from the original
TI-oscillatory to the TI-overdamped phase
by tuning the asymmetry parameter $\epsilon_{2x}$.

In FIG. \ref{gap_log} 
we show the thickness ($L_x$) dependence of the (half) size gap $E_0$
calculated by numerical diagonalization of the tight-binding Hamiltonian
with open boundary conditions at $x=0$ and $x=L_x +1$, 
superposed on
the same dependence predicted by Eqs. (\ref{deltaE2}).
The value of the parameter
$\epsilon_{2x}/m_{2x}$
is varied from its
original value
$\epsilon_{2x}/m_{2x} =0.11$
to $\epsilon_{2x}/m_{2x} = 0.3, 0.5, 0.7, 0.9$, which leads,
respectively, to the value of 
$\tilde{t}_x/m_{2x} = 1.57, 1.73, 2.10, 3.44$.
The corresponding point in the $(m_0/m_{2x}, \tilde{t}_x/m_{2x})$-plane
is specified in FIG. \ref{PD}, respectively, by
a filled square in purple ($\epsilon_{2x}/m_{2x} = 0.3$), a rhombus painted in light blue ($\epsilon_{2x}/m_{2x} = 0.5$),
 a upper green triangle ($\epsilon_{2x}/m_{2x} = 0.7$)
and a lower orange triangle ($\epsilon_{2x}/m_{2x} = 0.9$).
In FIG. \ref{gap_log}
the $L_x$-dependence of $E_0$
at these values of the parameter $\epsilon_{2x}/m_{2x}$
is indicated by points represented by the same symbol
and color.
The corresponding theoretical curve specified by
Eq. (\ref{deltaE2})
is superposed on the same figure
indicated by a continuous curve of the same color.
Since the phase boundary between
TI-oscillatory and TI-overdamped phases is located at 
$\tilde{t}_x/m_{2x} =1.8735$
on the line $m_0/m_{2x}= -1.3$,
$L_x$-dependence of $E_0$ also shows 
a crossover from
a damped oscillation (for $\epsilon_{2x}/m_{2x} =0.11, 0.3, 0.5$)
to overdamping (for $\epsilon_{2x}/m_{2x} =0.7, 0.9$).
This feature, together with an excellent agreement between
the numerical values of $E_0$ and Eq. (\ref{deltaE2}),
is highlighted in FIG.  \ref{gap_log}.

In FIG.  \ref{gap_log}
the size of energy gap determined experimentally for Bi$_2$Se$_3$ nanofilms
(reported in Refs. \cite{Hirahara}, \cite{exp_film1})
is also shown for comparison with our theoretical results.
It is curious that
those of Ref. \cite{exp_film1} (indicated in asterisks)
shows seemingly a monotonic decay as  a function of the thickness of the film,
while
those of Ref. \cite{Hirahara} (indicated in pluses)
seem to suggest an oscillatory behavior.
We also note that these experimental values are generally larger than
the theoretical predictions.

\section{Conclusions}

Motivated by experimental realization of the topological insulator thin films,
we have studied theoretically
possible size effects on the protected surface Dirac state on such films.
Indeed, finite thickness of the film
induces coupling between the two Dirac cones;
one at the top, the other at the bottom surface of the 
film ($=$ a slab-shaped sample of a 3D TI).
To quantify such finite-size effects we introduce
a 1D effective model of a topological insulator,
in which the spatial coordinate represents the direction of the depth,
{\it i.e.}, the direction vertical to the surface of the film.
Using this effective 1D model,
we reveal a precise correspondence between 
the thickness dependence of the size gap and
the spatial profile of the surface wave function
in the semi-infinite geometry.
In Sec. IV
we solved the boundary problem in the semi-infinite geometry.
This allows for, 
{\it e.g.}, determining exactly the energy at which the surface Dirac point appears.
We then use this solution to study
how the Dirac cone becomes gapped in the case of a slab with a finite thickness.

As shown in the phase diagram depicted in FIG. \ref{PD}
the topologically nontrivial phase of our 1D model Hamiltonian,
prescribed by Eqs. (\ref{h_1D}), (\ref{mass_1D}) and (\ref{epsilon_1D}),
is divided into two subregions:
the TI-oscillatory and the TI-overdamped phases.
This effective model is deduced 
from a more realistic 3D effective Hamiltonian
[specified by Eqs. (\ref{h_3D}), (\ref{mass}) and (\ref{epsilon})]
that successfully describes various topological features
observed, {\it e.g.}, on cleaved surfaces
of the Bi$_2$Se$_3$ crystal.
In such original 3D bulk TI,
asymmetry in the spectrum of valence and conduction bands
is omnipresent.
Here, we have demonstrated that by tuning this asymmetry
one can drive
[since tuning $\epsilon_{2x}$ leads to changing the effective hopping $\tilde{t}_x$
given in Eq. (\ref{tilde_t})]
a crossover from the TI-oscillatory to the TI-overdamped phase.
We have also established analytic formulas for the
the thickness ($L_x$) dependence of the size gap $2 E_0$
[Eq. (\ref{deltaE2})]
and a precise relation relation [Eq. (\ref{gap_wf})]
between
 $E_0 (L_x)$ and 
$\psi_{\rm semi} (x)$, 
the surface wave function in the semi-infinite geometry [Eqs. (\ref{psi_semi}), (\ref{sol_semi})].
Finally,
Eqs. (\ref{tilde_t}), (\ref{circle}), (\ref{deltaE2}) and (\ref{gap_wf}),
together with FIG. \ref{PD} and FIG. \ref{gap_log}
constitute the central results of the paper.

\acknowledgments
M.O. and K.I. thank useful discussion with Shuichi Murakami, Ai Yamakage and Takahiro Fukui.
K.I. also acknowledges Toru Hirahara, Koji Segawa, Yasuhiro Tada and Norio Kawakami for correspondences.
Y.T. is supported by Grants-in-Aid for Scientific Research (C) [No. 24540375] from Japan Society for the Promotion of Science.

\bibliography{1DTI_2}

\end{document}